# Highly multi-mode anti-resonant hollow core fibres

ROBBIE MEARS*, KERRIANNE HARRINGTON, WILLIAM J WADSWORTH, JAMES M STONE, TIM A BIRKS

*Centre for Photonics and Photonic Materials, Department of Physics, University of Bath, Bath, BA2 7AY, UK*
*rm2033@bath.ac.uk*

**Abstract:** We report the characterisation of anti-resonant hollow core optical fibres guiding at least 50 spatial modes in the infrared. Their propagation losses were measured to be between 0.1 and 0.2 dB/m from 1000 to 1500 nm wavelength, with bend losses of less than 3 dB/turn for bend radii of 7.5 cm despite core radii greater than 60 times the guided wavelengths.

## 1. Introduction

Anti-resonant hollow core fibres guide light through a gas or vacuum core. In this way the guided light is largely decoupled from the solid fibre material, greatly reducing material contributions to fibre non-linearity, damage thresholds and absorption [1,2]. This has enabled hollow core fibres to become the lowest attenuation fibres across a spectral range from the ultraviolet [3,4] to mid-infrared [5,6], surpassing solid core fibres even within the telecommunications bands [7].

Previous work on hollow core fibres have been focused on the design and fabrication of low-loss single-mode fibres for telecommunications applications [7-11]. This research has produced a wide range of single or few-moded fibres where the guidance is increasingly well understood [12-15], and kilometre-length fibres are now routinely reported [16]. In comparison, multi-mode hollow core fibres are relatively under-developed with remaining questions on their fabrication and optical performance. As a result, many applications which would benefit from the properties of a hollow core remain unaccompanied by a suitably multi-mode hollow core fibres, particularly at ultraviolet or mid-infrared wavelengths where solid core fibres exhibit prohibitively high loss.

A significant contributor to this imbalance in development is the general challenge of hollow core fibre fabrication. This requires the precise control of thin glass microstructures [17], and multi-mode hollow core fibres present different fabrication challenges from single-mode hollow core fibres. Consequently, many studies of multi-mode hollow core fibres have been theoretical only [18-20], with only a small number of fabricated fibres [21-24]. Within these few examples the number of guided modes has remained low, with a maximum of ~10 spatial modes reported [22,23], further limiting their deployment in highly multi-mode applications.

In this work we report the fabrication and characterisation of highly multi-mode anti-resonant hollow core fibres, designed to guide in the near-infrared wavelength range. Through a series of increasingly multi-mode fibres we follow the trend in modal content with capillary number and core size and demonstrate that hollow core fibres guiding up to ~50 spatial modes can be fabricated with low propagation losses and reasonable resistance to bend loss.

## 2. Design of multi-mode hollow core fibres

In anti-resonant hollow core fibres the guidance of light is based on the careful design and fabrication of thin glass capillaries, which confine light to a central core region through grazing incidence reflection. On each reflection some light is lost, leading to a fundamentally leaky guidance mechanism where higher-order modes experience a more rapid leakage of light due to steeper angles of reflection. By engineering the destructive (anti-resonant) interference of light within the thin walls of these capillaries their reflectivity is greatly enhanced, in turn reducing leakage losses. The destructive and constructive (resonant) interference conditions are well described by the ARROW model [25], which predicts the wavelength $\lambda_m$ of resonant bands

$$\lambda_m = \frac{2t}{m}(n^2 - 1)^{1/2} , \qquad (1)$$

where $t$ is the capillary wall thickness, $n$ the refractive index of silica and $m$ the resonance order. In an anti-resonant fibre this results in a series of high-loss resonances at integer values of $m$ ($m = 1, 2, 3, ...$) with wide low-loss ranges in between (around anti-resonances $m = 1/2, 3/2, 5/2, ...$). As a result, anti-resonant hollow core fibres can be designed and fabricated across a wide wavelength range through the simple variation of wall thickness.

The number of modes guided by a conventional (solid-core, step-index) multi-mode fibre is a straightforward concept because such a fibre has a well-defined cutoff condition, corresponding to the failure of total internal reflection at the core-cladding boundary. *For fixed cladding properties* (and a fixed wavelength), more modes are guided when the core size increases. In contrast, anti-resonant HCFs lack such a well-defined cutoff condition. Indeed, they are inherently multi-mode in principle, but higher-order modes suffer more leakage loss until (for a given fibre length) some are in-practice absent at the output. The number of guided modes therefore depends on the fibre's mode-dependent loss, as well as its length. Nevertheless, *for fixed cladding properties* (and a fixed wavelength), one might expect the effective number of modes guided by a particular length of HCF to increase with core size, as in a conventional fibre.

To study multi-mode HCF, we therefore fabricated a set of three fibres with different numbers of capillaries, but the capillaries had the same sizes, wall thicknesses and separations. This changes the core's radius simply because it changes its circumference. Since the optical properties of the cladding are largely determined by the capillaries, this is the closest we can get to achieving "fixed cladding properties". (Alternatively, the core's radius could be changed by fixing the number of capillaries but changing their sizes. However, this also changes their optical resonance properties, and so fails to achieve "fixed cladding properties". Indeed, HCFs with 6 closely-spaced simple capillaries are effectively single-mode regardless of the sizes of the capillaries and hence of the core [26].)

## 3. Experiment

To compare fibres with varying modal contents we fabricated three fibres with different numbers of capillaries using the stack and draw process, shown in Figure 1.

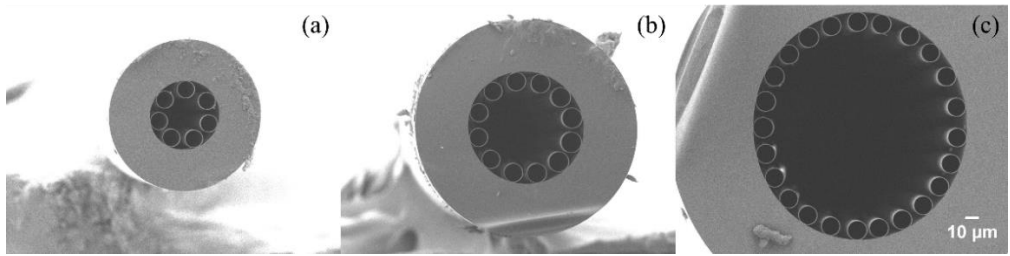

Figure 1. SEMs, to the same scale, of (a) the 7-capillary fibre (b) the 12-capillary fibre and (c) the 24-capillary fibre.

The fibres have 7, 12 and 24 capillaries and represent effectively single-mode, few-moded and many-moded fibres respectively. Other structural parameters of the three fibres are shown in Table 1, with capillary wall thicknesses inferred from optical transmission measurements using the ARROW model [25]; we find this to be more reliable than SEM measurements [3].

Table 1. Structural parameters for the three reported fibres.

| Number of capillaries | 7 | 12 | 24 |
|---|---|---|---|
| Inscribed core radius / µm | 12 | 27 | 67 |
| Capillary radius / µm | 6.5 | 6.8 | 6.5 |
| Wall thickness / nm | ~ 320 | ~ 370 | ~ 380 |
| Fibre outer diameter / µm | 110 | 160 | 280 |

While the fibres differ in their outer diameters, which is known to strongly affect micro-bending losses and mode coupling [13], they are similar enough in their capillary wall thicknesses and capillary radii to draw meaningful conclusions from their comparison. Based on Equation (1) the wall thickness range of 320 – 370 nm ensures significant overlap of the fundamental anti-resonant band (around $m = 1/2$), occurring for wavelengths above 900 nm for all three fibres.

For the basic characterisation of these fibres we used an incoherent white light source (Energetiq EQ-99X) with off-axis parabolic mirrors to couple the light into the fibre, providing a focused spot size of ~0.2 mm at the fibre input with a numerical aperture >0.2. The large numerical aperture ensured that a wide range of fibre modes (up to and beyond those expected to be guided) were uniformly excited. At the output the transmitted spectrum was measured by a Bentham DTMc300 double monochromator equipped with a range of detectors for wavelengths from the ultraviolet to the mid-infrared.

From the fibres' transmission spectra we can draw qualitative conclusions about their loss and modal content. Given the highly multi-moded incoherent source, the total power of the transmitted light is determined by a combination of fibre losses and number of guided modes. With each mode excited equally the total transmitted power scales with the effective number of guided modes [27]. The incoherence, high stability and large spot size of the source allows repeatable input coupling to a maximal value, permitting different fibres to be compared under similar coupling conditions. For our measured spectra, shown in Figure 2 for approximately 4 m of each fibre, the input coupling was optimized for maximum transmission around 1000 nm in each case.

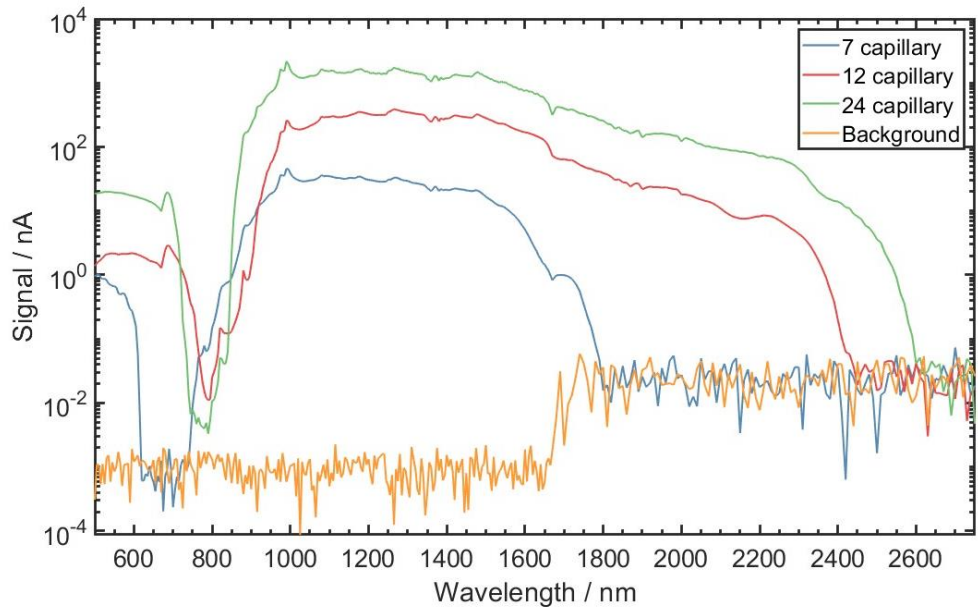

Figure 2. Transmitted power (measured photodiode current) against wavelength on a logarithmic scale for ~ 4 m of each fibre, with a background scan taken by blocking the light source. The step change in measured background at 1700 nm is due to a change in monochromator detector.

In Figure 2, and later analysis, we focus on the fundamental band ($m < 1$), with the first high-loss resonance ($m = 1$) occurring between 700 – 800 nm for all three fibres. Away from this resonance the total transmitted power increases markedly with the number of capillaries, consistent with a greater number of modes. In the relatively flat range of the spectrum around 1250 nm, the 12 capillary fibre and 24 capillary fibres respectively transmit ~11 × and ~50 × more light than the 7 capillary fibre. Based on near field images and bending we observed that the 7 capillary fibre transmits a single fundamental mode along with some higher order mode content which were unable to quantify. Nevertheless, if we assume the transmitted power for similar overfilled incoherent input conditions is proportional to the number of modes, we can conclude that the 12 and 24 capillary fibres guide *at least* 11 and 50 spatial modes (neglecting two-fold polarisation degeneracy) respectively. (If the 7 capillary fibre guides more than one mode then the other fibres must guide proportionally more modes.) These transmission measurements provides a quick estimate for the number of guided modes, using an exceedingly simple experimental setup. A more quantitative analysis may be achieved through an $S^2$ experiment [28], which measures the number of modes based on their differential group delay. This would be likely to be a challenging experiment due to the large number of spatial modes and the practical need to match fibre length and spectrometer performance to the expected group delays, with previous $S^2$ experiments only identifying up to ~10 spatial modes [22].

Another interesting trend is the extension of the fundamental low-loss band towards longer wavelengths in fibres with more capillaries. The 12 and 24 capillary fibres show greater transmission at longer wavelengths, widening the effective bandwidth. Furthermore, the approximate doubling of core size from 12 to 24 capillaries does not yield the same bandwidth extension as the jump from 7 to 12 capillaries. We attribute this behaviour to the interaction of guided core modes with glass capillary modes, which is typically avoided by the large difference in spatial frequencies between the two sets of modes [11]. This coupling is often seen at the red edge of anti-resonant bands [11,14,29] and effectively curtails the bandwidth of the fundamental band. In our series of fibres the glassy modes are approximately the same across the different fibres (given the similar capillary size and thickness), while the increase in core size causes a given core mode to vary less rapidly relative to the capillary glassy modes. In a coupled mode picture this would result in less interaction with glassy modes, extending the red edge of the fundamental band.

For the measurement of fibre attenuation, we performed cutback experiments using the same setup with each fibre laid in loose loops of ~0.5 m radius to minimise any bend loss. The fibre lengths were 30 – 50 m before and ~10 m after cutback in each case, with the resulting attenuation spectra shown in Figure 3. In multi-mode fibres, the measured attenuation will depend strongly on the cutback conditions such as length, modal excitation and the degree of inter-mode coupling. These relatively-short cutback lengths will yield higher attenuation values than longer cutbacks would because they include lossy higher-order modes. However, it presents a realistic attenuation value for their use in short length applications where highly multi-mode fibres are most likely to be employed.

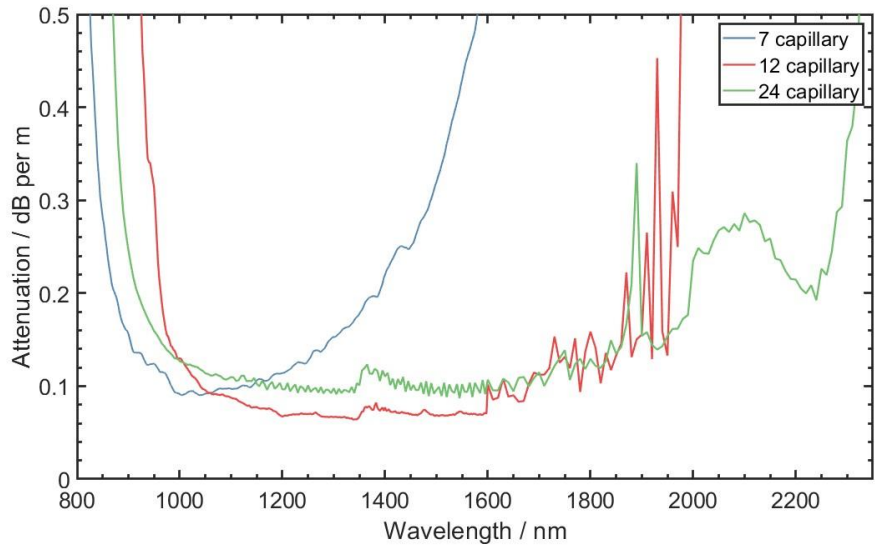

Figure 3. Attenuation plots of the three fibres calculated from separate cutback experiments. The noise in the curve for the 12 capillary fibre beyond 1800 nm is due to detector noise combined with a lower signal level compared to the 24 capillary fibre.

The cutback measurements display a similar minimum attenuation across all three fibres, hovering around the 0.1 dB/m level. As was noted previously, the fibres with more capillaries present a wider low-loss bandwidth, with the 24 capillary fibre extending from 900 to 2000 nm with attenuation below 0.2 dB/m. This represents a significant extension in low-loss bandwidth when compared to the 7 capillary fibre which only spans from 850 to 1400 nm below the same attenuation level. Wider bandwidths in HCFs are highly desirable, particularly for applications such as gas sensing, which could enable broader spectral detection of a wider range of molecules within a single fibre.

Finally, we measured macro bend losses by comparing the transmitted signal through a loosely straight ~4 m section of fibre to that through the same section bent around discs of different radii. The discs were placed approximately half way along the fibre length, with roughly straight sections leading into and out of the loop. Due to higher bend losses from the larger cores, we measured only a single loop around each disc with the resulting bend loss spectra displayed in Figure 4.

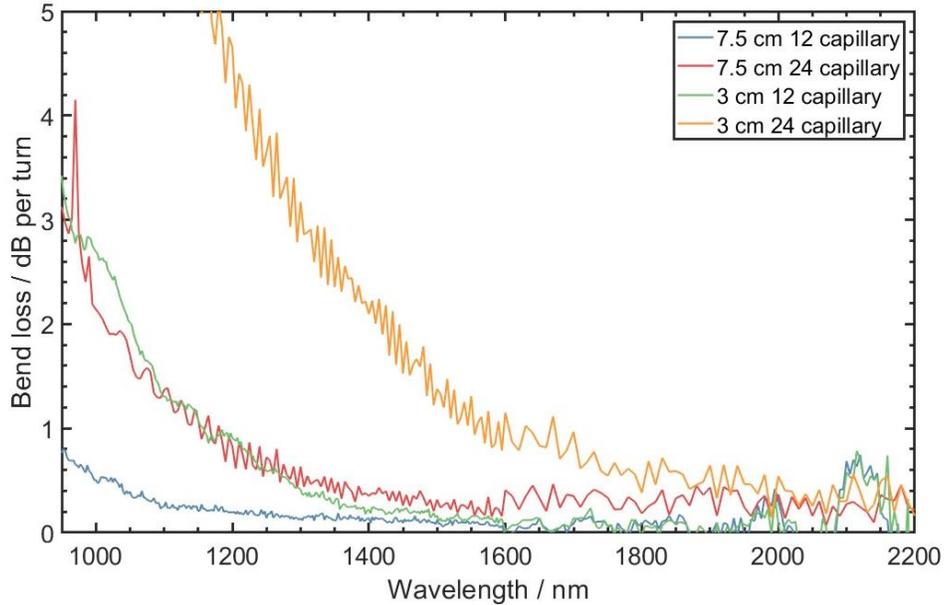

Figure 4. Bend loss spectra for a single turn around discs of the indicated radii. The 7-capillary fibre is not shown.

The bend loss of the 7 capillary fibre was too small, <0.1 dB/turn across this spectral range, to show in the figure. The larger core sizes in the 12 and 24 capillary fibres results in greater bend losses, with the largest core demonstrating the highest bend induced loss. Nevertheless, for a moderate bend radius of 7.5 cm, both fibres display bend losses < 1 dB/turn for much of their fundamental band but this rises to the 2 – 3 dB/turn level for the tighter bend of 3 cm radius. Given that many applications do not require such tight bends, and fibres are often stored on standard 15 cm diameter shipping spools and used with bends of similar diameter, these bend losses may be acceptable for a range of uses where a higher number of modes is desired. As the fibre length before the bend was short and as we used a large area incoherent excitation, we expect some high-order modes to be propagating in the straight fibre. These modes will be most susceptible to bend loss, which could contribute to the relatively high bend loss seen.

While the bend induced losses preclude their use in applications requiring many tight loops they are better than one might first expect for fibres with core diameters of 100 or more times the wavelength. If we consider that macro bend loss is correctly described by resonant bend induced coupling between core and capillary air modes [30,31], increasing the difference in size between core and capillary regions should make the fundamental mode more resistant to bend loss. In practice, the presence of higher-order modes will offset this advantage as they become increasingly sensitive to macro bending as mode order increases. The measured bend loss is thus a combination of the different modes and will further depend on how mode coupling transfers power between modes. While a quantitative analysis of mode coupling is beyond the scope of this paper, the moderate bend losses for large core diameters would suggest such mode coupling is not dominant over ~4 m lengths with bending halfway along the length.

### 4. Discussion

In our series of fibres we investigated multimode properties of hollow core fibres by increasing the core diameter while maintaining roughly the same capillary diameter and spacing. This approach is advantageous over holding the core diameter constant, because it maintains the optimum grazing incidence angle, resonant coupling to cladding capillaries, and azimuthal confinement for high order modes in the multimode fibre as in the equivalent single mode fibre. This allowed the guidance of ~50 spatial modes while maintaining reasonable bend resistance.

By holding the wall thickness constant, we additionally probed the effect of varying the core diameter for a fixed wall thickness, with significant bandwidth improvements for larger core diameters. As the bandwidth of these fibres is key to their operation and use, the ability to enhance a fibre bandwidth by adding capillaries is a useful tool for hollow core fibre design and fabrication. In contrast, doubling the core size for a given wall thickness without adding capillaries would require the capillary thickness to diameter ratio to become twice as small, leading to an unstable fabrication process [17]. If the same result is achieved by adding capillaries, the fabrication challenge is in the placement of additional capillaries, without any change in the capillary aspect ratio.

While the reported fibres are highly multi-mode in comparison to other reported hollow core fibres, a typical solid core fibre with a similar size core at these wavelengths would undoubtedly guide hundreds of modes. The benefit of a multi-mode hollow core is the suppression of material contributions to non-linearity, dispersion and absorption which can enable certain applications requiring ~10 – 100 spatial modes [32]. These advantages become particularly large in the ultraviolet or mid-infrared, where high power laser sources are commonly multimode and comparable solid core fibres are not currently competitive due to inherent material absorption losses.

## 5. Conclusion

We report a series of anti-resonant hollow core fibres with increasing modal content. These fibres were carefully designed and fabricated with similar capillary size and wall thickness, showing that larger core diameters yield greater numbers of guided modes and wider low loss bandwidths. Despite their large core diameters, reasonable bend resistance was observed which suggests that inter-mode coupling was not dominant. By comparing their transmission of light from an incoherent multi-mode source, we estimated the number of spatial modes in our 24 capillary fibre to be at least 50, representing a nearly 5-fold improvement in modal capacity over previously-reported hollow core fibres.

**Acknowledgements**

The authors would like to thank David Bird and Jonathan Knight for useful discussions, and comments on the manuscript.